\documentclass[%
 reprint,
superscriptaddress,
 amsmath,amssymb,
 aps,
prm,
]{revtex4-2}

\usepackage{graphicx}% Include figure files
\usepackage{epstopdf}
\usepackage{dcolumn}% Align table columns on decimal point
\usepackage{bm}% bold math
\usepackage{gensymb}
\usepackage{upgreek}
\usepackage{hyperref}
\usepackage{tabularx}
\usepackage{newunicodechar}
\newunicodechar{−}{\textminus}

\begin{document}
\author{Michelle E. Jamer}
 \email{jamer@usna.edu}
\affiliation{Physics Department, United States Naval Academy, Annapolis, MD 21402, USA}
\author{Gregory M. Stephen}
\affiliation{Laboratory for Physical Sciences, College Park, MD 20740, USA}
%\author{Adrian Fedoroko}
%\affiliation{Physics Department, Northeastern University, Boston, MA 02115, USA}
%\affiliation{Plasma Science and Fusion Center, MIT, Cambridge, MA 02139, USA}
\author{Brandon Wilfong}
\affiliation{Physics Department, United States Naval Academy, Annapolis, MD 21402, USA}
\author{Radhika Barua}
\affiliation{Mechanical and Nuclear Engineering, Virginia Commonwealth University, Richmond, VA 23220, USA}
\author{Frank M. Abel}
\affiliation{Physics Department, United States Naval Academy, Annapolis, MD 21402, USA}
\author{Steven P. Bennett}
\affiliation{Materials Science and Technology Division, U.S. Naval Research Laboratory, Washington, District of Columbia 20375, USA}
\author{Joseph C. Prestigiacomo}
\affiliation{Materials Science and Technology Division, U.S. Naval Research Laboratory, Washington, District of Columbia 20375, USA}
\author{Don Heiman}
\affiliation{Physics Department, Northeastern University, Boston, MA 02115, USA}
\affiliation{Plasma Science and Fusion Center, MIT, Cambridge, MA 02139, USA}
\author{Dave Graf}
\affiliation{National High Magnetic Field Laboratory, Florida State University, Tallahassee, FL 32306, USA}

\date{\today}

\begin{abstract}

Fe$_3$Ga$_4$ possesses a helical spin spiral with a complex competition between ferromagnetic and antiferromagnetic ground states. This competition generates multiple metamagnetic transitions that are governed by both applied magnetic field and temperature. At intermediate temperatures between T$_1$ (68 K) and T$_2$ (360 K), the ferromagnetically aligned spins transition to an antiferromagnetic spin spiral. In this study, magnetoresistance (MR) measurements are performed on an aligned single crystal and compared to magnetization properties in order to gain insight on the unique alignment of the spins. The high-field MR is positive at low temperatures indicating cyclotronic behavior and negative at high temperature from electron-magnon scattering. Of particular significance is a large anomalous positive MR at low fields, possibly due to emergent spin fluctuations thus prompting further exploration of this multifaceted material. 

\end{abstract}

\title{Anomalous low-field magnetoresistance in Fe$_3$Ga$_4$ single crystals}
%\title{Elucidating the spin structure of Fe$_3$Ga$_4$ with transport properties}

\maketitle
\begin{figure*}[]
    \centering
    \includegraphics[width = 0.98\textwidth]{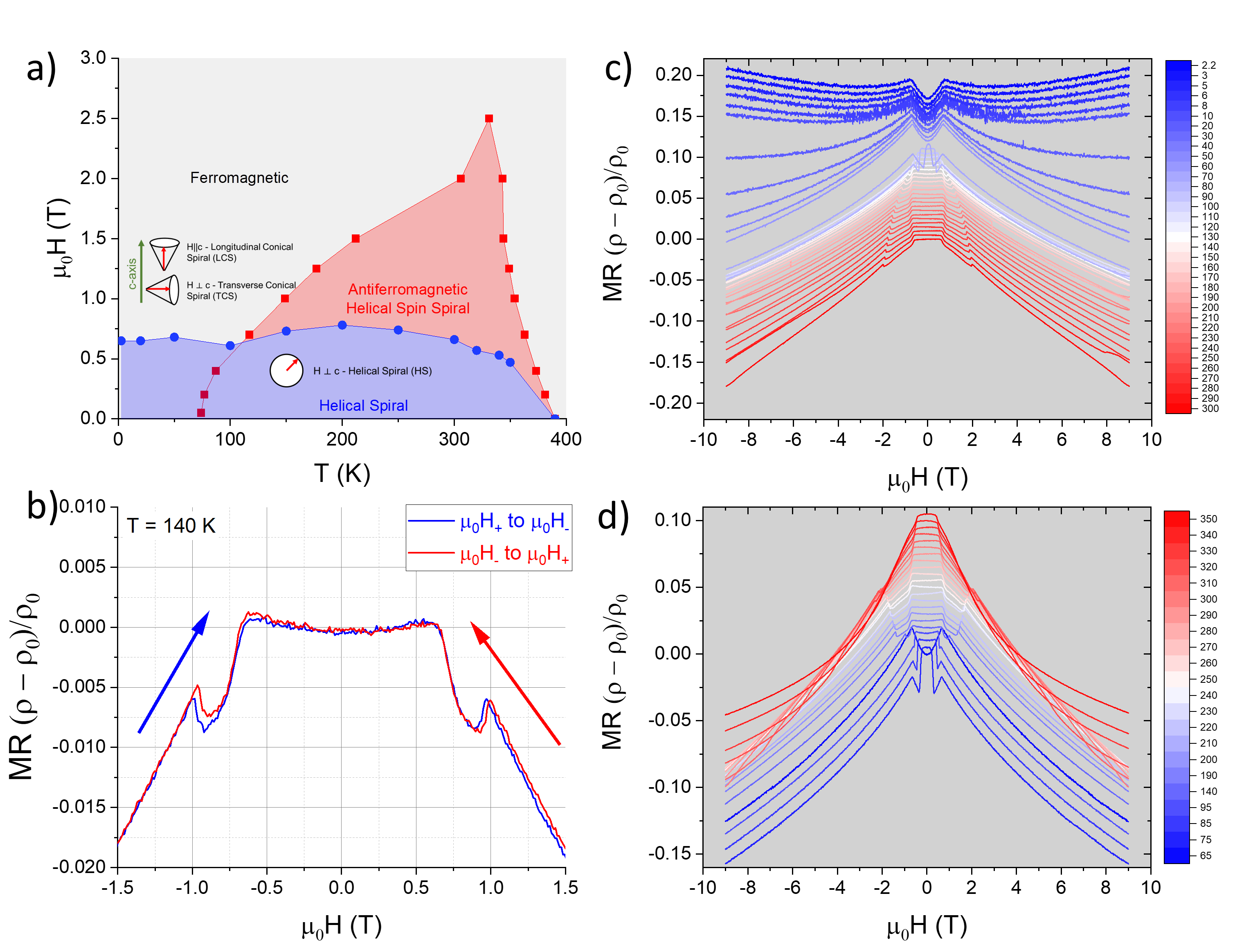}
     \caption{(a) The magnetic phase diagram with applied field ($\mu_0$H) and temperature. The antiferromagnetic state is shown in red, where the red region is given by the change in the magnetic transitions T$_1$ and T$_2$. The values noted in the diagram are from the $b$-axis as noted in the supplemental information.\cite{SM} The helical spiral (HS) in the diagram in blue is the magnetic step along the $b$-axis, where the low-field behavior is a two dimensional spin spiral,. The HS is also observed along the $a$-axis in the antiferromagnetic region as noted in the supplemental information.\cite{SM} (b) The magnetoresistance (MR) along the $b$-axis at T = 140 K, where the crystal is in the antiferromagnetic state. The down sweep is in red and the up sweep is in blue. There is a hysteresis oberved at $\mu_0$H $\approx$ $\pm$ 1.0 T, which relates to saturation at T = 140 K. There is a smaller change close to $\approx$ 0.75 T, which is the field that the HS transitions to the TCS. (c) The MR at each temperature along the $b$-axis starting from 2.2 K to 300 K. (d) The MR at each temperature along the $b$-axis from 350 K to 65 K.}
    \label{fig:spiral}
\end{figure*}
There is a crucial need to explore magnetic materials with novel magnetic transitions in order to keep pace with the demands of an increasingly digital society.\cite{Review3} One material of interest has been FeRh for a variety of studies due to its antiferromagnetic (AFM) to ferromagnetic (FM) transition near room temperature. AFM-to-FM transitions are rather rare in materials, and FeRh has been proposed for a variety of applications, including devices that use magnetization as a state variable and logic.\cite{FeRhReview, FeRhReview2, Barua, Bennett2016} While this compound has been used as a test bed for possible devices,\cite{Blumenschein, FeRhDevice1} FeRh is hindered by the expensive Rh component, the structural transition accompanying the magnetic transition that can increase brittleness, and the sensitivity of the transition temperature to material quality.

%In addition to the use of magnetic-induced transitions in devices, there are as yet unknown possibilities for memory applications relying on magnetic skyrmions.

Fe$_3$Ga$_4$ has recently been explored as a compound similar to FeRh due to its metamagnetic transitions, but Fe$_3$Ga$_4$ has the additional feature of a helical spin spiral (HSS) phase.\cite{Kawamiya,Mendez,WilfongSC,AHEFe3Ga4} Like FeRh, Fe$_3$Ga$_4$ is metallic with a metamagnetic transition near room temperature ($\approx$ 360 K) but without the accompanying structural transitions. The compound forms in the monoclinic C2/m structure with four unique Fe atoms, which allows for the helical spirals with crystallographic information in the Supplemental Material (SM).\cite{SM} Fe$_3$Ga$_4$ has two magnetic transitions of interest, which are both sensitive to applied magnetic field and temperature. At T$_1$=68~K, the system transitions from a low-temperature FM state to a helical spin spiral (HSS) AFM state, while at T$_2$= 360 K the system transitions from  the HSS AFM back to the FM state as illustrated in Fig. \ref{fig:spiral}(a). The magnetic transitions are caused by a sensitive competition between ground state energies of the FM and  HSS AFM structures, where small energy changes cause a change in the magnetic structure. 
%Due to T$_2$'s proximity to room temperature, and the fact that there is no accompanying structural transitions to the change in the magnetism, Fe$_3$Ga$_4$ could be useful in a variety of potential magnetic devices in lieu of FeRh.

Some original studies of Fe$_3$Ga$_4$ focused on the metamagnetic transition T$_1$ and the effects of chemical dopants.\cite{Wilfong2022, Kawamiya} Recently, single crystal studies have been aimed at understanding the configuration of the spin system.\cite{Mendez,WilfongSC,Afshar} Part of the renewed interest is due to the identification of a topological component in the Hall effect measurements.\cite{Wu,WilfongSC} In these newer studies, the magnetic and resistivity properties have been studied.\cite{baral2025fluctuation} The magnetic properties along each crystallographic direction demonstrate that there is a HSS-AFM state sandwiched between two FM states. The magnetic transitions vary as a function of both field and temperature, as visualized in the magnetic phase diagram in Fig. \ref{fig:spiral}(a) with full information regarding transition temperatures and fields in the Supplemental Information. As seen in the magnetic phase diagram, along the $a-$ and $b-$axis, the magnetization behaves as a two-dimensional helical spiral in the low-field range ($\mu_0H \leq 0.75$ T), then transitions into conical structures at higher fields.

%The change in the magnetization with respect to temperature also gives insights in the magnetic stepping behavior in the magnetization measurements. As seen in Fig. \ref{fig:spiral}(d), there are various steps where the magnetization changes based on the interaction between the applied field and spiral. Along the $a$- and $b$- axis, the magnetization behaves as a two-dimensional helical spin spiral in the low field range. In Fig. \ref{fig:spiral}(d), the applied magnetic field where the first magnetic step occurs is determined to be linear with respect to temperature. It is noted that the $b$-axis always has a magnetic step when $\mu_0 H \approx 0.75$ T. The graph shows the following step after that transition field. Furthermore, as the temperature increases, the field where the sample saturates increases as illustrated in Fig. \ref{fig:spiral}(e). 

In this work, magnetoresistance (MR) measurements are used in order to understand the coupling between the electrical transport behavior and the magnetic phase properties. These measurements can give more insight on the magnetic coupling in the HSS-AFM state as well as how these states could be utilized in a practical device. While many AFM-based states are insulators, the spin spiral ordering allows for retained metallicity at room temperature. In this paper, we investigate the crystallographic axes-dependent transport and magnetotransport properties, leading to a deeper understanding of the spin structure. Of particular curiosity is the MR ($MR = \frac{\rho_H - \rho_0}{\rho_0} $) along the $b$-axis, which has a hysteretic MR in the AFM region as the spin structure changes from helical spiral (HS) to transverse conical spiral (TCS) (Fig.\ref{fig:spiral}(b)) and a large anomalous positive MR (pMR) at low magnetic fields that persists at room temperature as seen in Fig. \ref{fig:spiral}(c) and (d). The anomalous pMR at low fields correlates to a large electron mobility which is typically only identified at high fields for FM materials can indicate helical ordering at low fields.

\begin{figure*}
    \centering
    \includegraphics[width =0.85\textwidth]{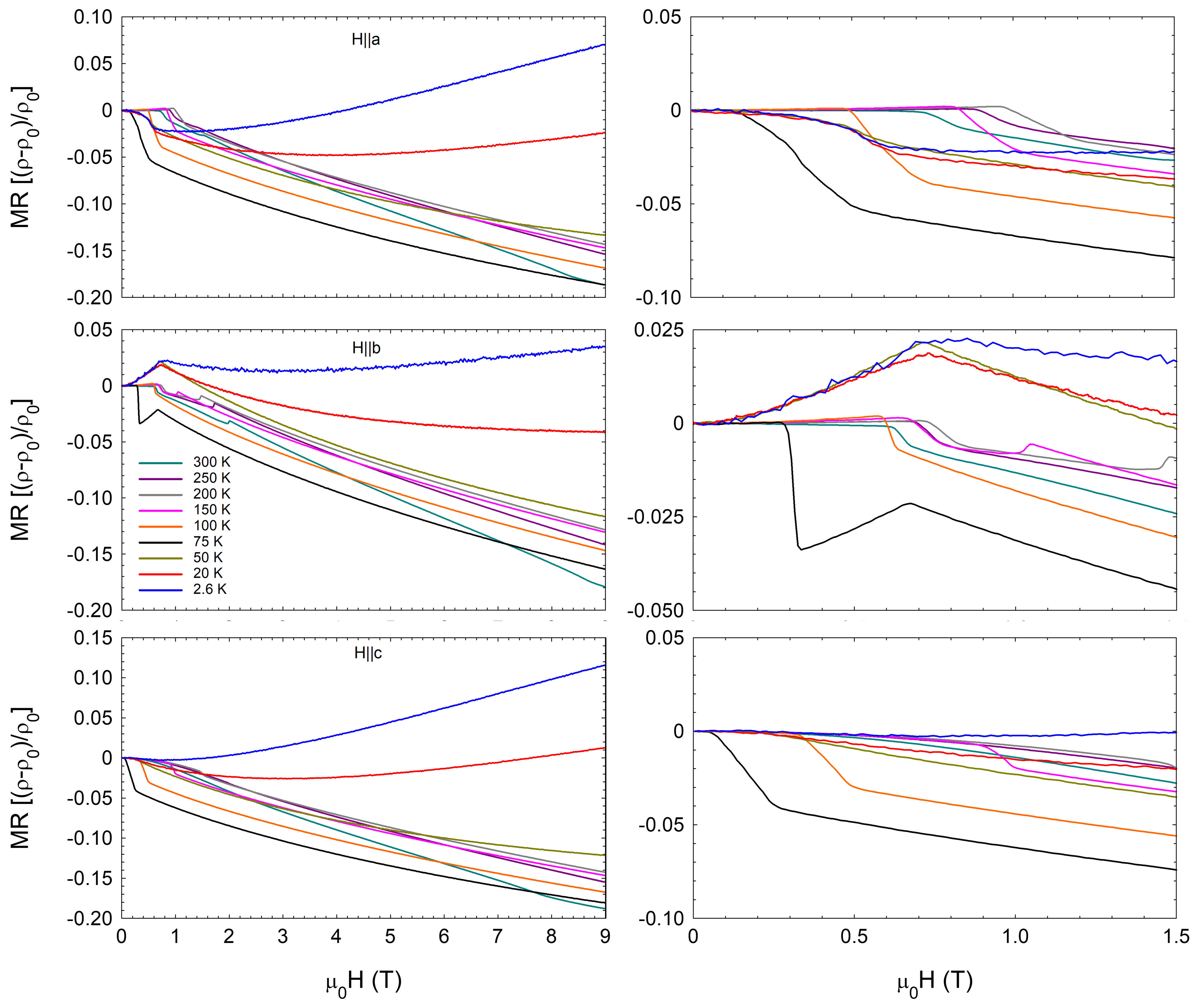}
    \caption{(Left) The data for the magnetoresistance (MR) when $\mu_0 H $ is applied along the $a$, $b$, and $c$ crystallographic axes. (Right) The same data as the (left), but focused on the region when $\mu_0 H = 0 - 1.5$ T. The data shows that the MR along the $b-$ axis has a persistant positive contribution below 0.75 T for all temperatures with the exception of the 300 K data. The $a$-axis has a slight positive magnetoresistance for temperatures when the crystal is in its Helical Spin Spiral (HSS) state after T$_1$. The legend for temperatures correspond to all curves in this figure.   }
    \label{fig:mr}
\end{figure*}
\begin{figure*}
    \centering
    \includegraphics[width =0.95\textwidth]{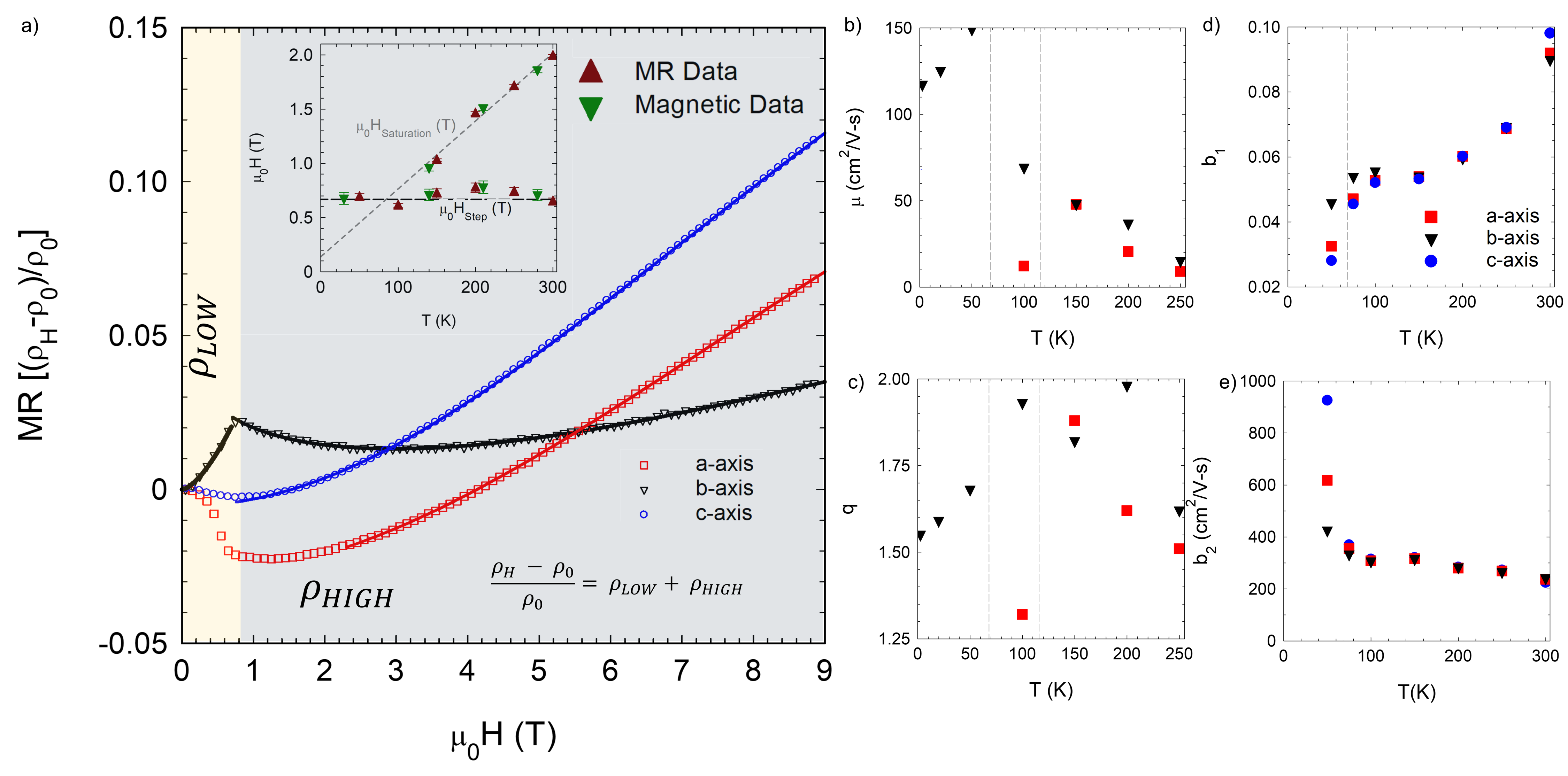}
    \caption{(a) An example of the fits for the $a$, $b$, and $c$-axes at 2.6 K as noted by solid lines, and the regions for designated field ranges are noted by different colors ($\rho_{low}$ is light yellow and $rho_{high}$ is light grey). (inset) The field where saturation occurs and the magnetic step indicating a change from the HS state to the transverse conical spiral (TCS) state along the $b$-axis. Magnetic data describes the data extrapolated from SQUID magnetometry in reference \cite{WilfongSC} and MR data is the field where the MR decreases in the intermediate field and temperature range. (b) and (c) The fitting values from modeling the low-field positive magnetoresistance where $MR = (\mu \mu_0 H)^q$, where $1<q<2$. The positive magnetoresistance is only found along the $b$-axis.(d) and (e) The values from modeling the high-field data that was fit to the electron-magnon model where MR = $-b_1 \ln(1 + b_2^2 (\mu_0H)^2)$ for the $a$, $b$ and $c$ axes.  }
    \label{fig:constants}
\end{figure*}
%Elemental Fe granules (Alfa Aesar 99.98 \%) and Ga solid (Alfa Aesar 99.99 \%) were added in a molar ratio of Fe:Ga (68:32) to a quartz tube along with I$_{2}$ chips (Alfa Aesar 99.5 \%) in a concentration of 18 $\times 10^{-6}$ moles/cm$^{3}$ based on the approximated volume of the quartz tube. The elements inside the quartz tube were then evacuated to $<$ 7 $\times 10^{-3}$ Torr and sealed. The quartz tube was arranged in a triple zone furnace with the charge situated in the hot-end set to 700 $^{\circ}$C and such that the cold-end was set to 650 $^{\circ}$C in the remaining furnace zones. The growth was held at these temperature settings for four weeks before the furnace was turned off and allowed to cool to room temperature naturally. Long needle-like single crystals (approximately 1 mm wide and 0.25 mm thick with variable lengths up to 2 cm) were recovered from the cold-end of the quartz tube and washed with methanol. The structural and magnetic properties of the recovered Fe$_{3}$Ga$_{4}$ were previously confirmed

Single crystals of Fe$_{3}$Ga$_{4}$ were synthesized using a chemical vapor transport (CVT) method which was adapted from previous work. \cite{Philippe} Detailed synthesis methods and preliminary properties are described in a prior publication.\cite{WilfongSC} The transport properties were measured at the National High Magnetic Field Laboratory (NHMFL) DC Field Facility. The MR data was taken between 0 and 9 T, and from 2.2 to 300 K and seen in Fig. \ref{fig:mr}. 

The MR data differs when the field ($\mu_0H$) is applied parallel to the $a$-, $b$- and $c$- axes, as seen in Fig. \ref{fig:mr}. The low-field MR data for all crystallographic axes are highlighted in Fig. \ref{fig:mr}(right). Most interesting is the $b$-axis data where there is a strong anomalous pMR at low fields and low temperatures, but remains small and persistent even up to room temperature. Interestingly, around the metamagnetic transition (T$\approx$ 75 K), there is no pMR in the low field region. The MR data is modeled in two magnetic field regions which are labeled as $\rho_{low}$ and $\rho_{high}$. The entire field and temperature range can be fit using the following expression

\begin{equation}
    \frac{\rho_H-\rho_0}{\rho_0} = \rho_{low} + \rho_{high}
\end{equation}

and examples of the fit is seen in Fig.\ref{fig:constants}(a). 

The low-field component $\rho_{low}$ is a region of anomalous pMR, which is large at low temperatures (T $\leq 75$ K) along the $b-$axis. The low-field component corresponds to the field range $\mu_0 H \leq 0.75$ T, where the $a-$ and $b-$ axes are in the HS spiral phase as seen in Fig. \ref{fig:spiral}(b). The $\rho_{low}$ is large in the $b-$axis for all temperatures below 75 K, but is seen for all temperatures below 300 K with the exception of 75 K. Along the $a-$axis, a pMR emerges for the temperature range $100~K \leq T \leq 250~K$. The low-field data described by $\rho_{low}$ allows for more insight on the magnetic structure of Fe$_3$Ga$_4$. The MR in this field range highlighted in Fig. \ref{fig:mr}(right) indicates that the $c$-axis data has no positive magnetoresistance at temperatures measured with magnetic field below 0.75 T. We note that pMR at low fields is a characteristic of AFM materials, and we used the empirical equation
\begin{equation}
    \rho_{low} = (\mu \mu_0H)^q
\end{equation}
where $1 < q <2$ and $\mu$ correlates to the mobility of the carriers.\cite{PhysRevB.90.024403,saha,Nagasawa1977} Interestingly, the behavior fits to a semi-quadratic model, which is associated with s-d scattering.\cite{Yamada1973,Nagasawa1972,Azarevich_2022} There, it was pointed out when the field is parallel to the spins of one of the sublattice, the fluctuations in that sublattice may be suppressed, whereas the fluctuations in the other sublattice may be increased. A linear pMR in AFMs and density wave materials, such as Cr \cite{Arajs1965} and 2H-NbSe$_2$\cite{Naito1982}, is a general phenomena arising from partially gapped Fermi surfaces\cite{Feng2019}. Since the effect is seen below 0.75 T, when the magnetization is in the HS state, the positive MR is most likely due to spin fluctuations as proposed by the for Rivier and Zlatic model.\cite{Zlatic1978, Zlatic1981} A normal ferromagnet at 2 K has a relatively low mobility of $< 50 \frac{cm^2}{Vs}$ and ferromagnets with interesting topological properties normally have large mobilities with large MR as in the case of MnBi which has a large mobility of $5,000 \frac{cm^2}{Vs}$ and a positive MR of 250\% at 2 K. \cite{MnBi} The mobility of Fe$_3$Ga$_4$ in its low-field and low-temperature state has a mobility that is approximately 30 times larger than a regular ferromagnet at 50 K with a mobility of $\approx$ $150  \frac{cm^2}{Vs}$ at low fields with a small positive MR. This is particularly interesting since topological and helical materials have a positive MR with a large mobility due to smaller effective masses.\cite{Lee,Kanazawa,Nagaosa} Due to the size of the mobility and the anomalous pMR at low fields, we expect that the state that has traditionally been attributed to traditional FM order may have helical processions. 

The data for the fit for $\rho_{low}$ is seen in Fig. \ref{fig:constants}(b) and (c). In the figure, two temperatures are noted by a grey line which note where $T_1$ is located in the metamagnetic transition with no applied field (68 K) and with 0.75 T (116 K). The data for 75 K is not noted since near the transition temperature, there is no positive component, since this temperature correlates to the metamagnetic transition. The data below 68 K increases linearly for the $\mu$ and $q$ fitting factor along the $b$-axis, and then decreases. When $T \geq 100 K$, the $a-$axis has a similar spin structure to the $b-$axis. The low-field MR along the $a-$axis moves from negative to positive, where the mobility is 10 times smaller than along the $b$-axis. The disparity emerges because the metamagnetic transition T$_1$ occurs near 100 K. However, the data for the $a$-axis matches well after this temperature (well beyond the transition temperature) for both the mobility $\mu$ and the exponent $q$. Since both the $a$- and $b$-axes both demonstrate helical spiral ordering at low fields beyond the metamagnetic transition as noted by Fig.\ref{fig:spiral}(a), it is reasonable that their fitted values would be similar in the temperature region $150 \leq T \leq 250$ K. There is no positive low-field MR at room temperature (300 K) where the compound is moving towards its ferromagnetic alignment with applied field.

Along the $b-$ axis, there is a notable decrease in the MR between $0.75 \leq \mu_0H \leq 3$ T when T$\geq$ 75 K. When measuring above the HS to TCS transition (T$_1$ $\approx$ 70 K) there is a step-like behavior where the MR becomes more negative. The magnetic field step and magnetic saturation along the $b$-axis are both noted in Fig. \ref{fig:constants}(a,inset). The field where the decrease in resistance starts is noted by $\mu_0H_{Step}$, and is $\approx$ 0.75 T for all temperatures.  The resistance increases again when the magnetic state changes from the transverse conical spiral to the ferromagnetic state when the field saturates the compound, noted by $\mu_0H_{Saturation}$ in Fig. \ref{fig:constants}(a,inset). Interestingly, the main hysteresis locations observed in Fig. \ref{fig:spiral}(b) from field up from -9 T (blue) and down from +9 T (red) correlate to the magnetic ordering change from HS to TCS at 0.75 T. The second hysteresis field location occurs at $\pm$ 1.0 T, which is the saturation field at that temperature. After saturation, the MR behaves in the traditional electron-magnon model and matches well with the MR data from $a-$ and $c-$axis with a slight offset, as seen in the SM.\cite{SM} The decrease in resistivity along the $b$-axis is directly correlated to the TCS state, which persists to slightly past room temperature and is fully extinguished at 350 K, as seen in Fig. \ref{fig:spiral}(d). There is a similar decrease in resistance along the $a$- and $c$- axis when the metamagnetic transition occurs ($\mu_0 H \geq 0.75 T$), which is ascribed to a change in the magnetization from the FM-type state to another spin spiral state as seen in Fig. \ref{fig:spiral}(b).

The component $\rho_{high}$ corresponds to the applied magnetic field above $\mu_0 H \geq 0.75T$ T. The high-field data in all three crystallographic directions can be fit to the Khosla-Fischer model which describes electron-magnon scattering.\cite{Marrows2} The model describes the MR in systems with localized magnetic moments

\begin{equation}
\rho_{high} = -b_1 \ln(1 + b_2^2 (\mu_0H)^2) + \frac{b_3^2(\mu_0H)^2}{(1 + b_4^2(\mu_0H)^2)}
\end{equation}

, where the first term is the negative MR from electron-magnon scattering and the second term is positive MR from orbital cyclotronic behavior. Below 50 K, there is a positive MR at high fields due to an orbital component that behaves semi-quadratically.\cite{PhysRevB.90.024403}. The value of the quadratic orbital cyclotronic behavior varies empirically, and is corrected using a term in the denominator to fit the experimental data. The denominator term ($1 + b_4^2(\mu_0H)^2)$ was originally proposed by Sondheimer and Wilson as a two-band model.\cite{Sondheimer} The values of each fitted value in all models are noted in the Supplemental Information, and $b_1$ and $b_2$ are noted in Fig.\ref{fig:constants}(c) and (d), respectively. The values for $b_2$, $b_3$, and $b_4$ correlate to the mobility, and are proportional to the value of field. These values are notably higher than expected for a ferromagnetic material at low temperatures, which indicates unique topology states.\cite{MnBi} At 50 K, the mobility noted by $b_2$ is close to 1000 cm$^2$/(Vs) along the $c$-axis which has a MR of -12\% at 9 T and 50 K. When used for the high field data along the $b$-axis for $T < 75 K$, the pMR for this model normally resulting from an orbital cyclotronic component would give an unreasonable value for the mobility of 10$^5$ $\frac{cm^2}{Vs}$, and not considered further. We believe that the pMR along the $b-$axis is best described by the Rivier-Zlatic model, which describes the effect of localized spin fluctuations, whereas the Khosla-Fischer model proposes a similar quadratic fit from a third order perturbation of the exchange Hamiltonian.\cite{Zlatic1978, Zlatic1981}

The electron-magnon model correlating to $\rho = -b_1 \ln(1+b_2^2(\mu_0H)^2)$ is used solely for the data when T $\geq 50$ K. The electron-magnon model includes a nonsaturating linear magnetoresistance when the applied field is larger that the saturation field of the structure. The metamagnetic material's properties behave as a ferromagnet in this field range, and the fitting parameters $b_1$ and $b_2$ are essentially the same when $T > T_1$ for all three crystallographic directions, which is marked by a grey line in the figure. The $b_2$ parameter which correlates to the mobility of the carriers decreases exponentially through the metamagnetic transition, but then stabilizes to a constant value when the compound is in its helical spin spiral state for each crystallographic direction. 

The MR measurements support the results in the magnetization measurements and the large mobility indicates a unique topology of the structure.\cite{WilfongSC} In previous studies, a similar positive MR at low fields below a critical field was attributed to an electron-spin scattering of the {\it{s-d}} configurations in an antiferromagnetic metal.\cite{Yamada1973} When the field is parallel to the magnetization in the AFM state, it induces a positive MR below the critical field which is around 0.75 T.\cite{Nagasawa1972} The intermediate field can be considered the intermediate state below a secondary critical field, which changes with temperature due to the metamagnetic properties.\cite{Yamada1973} After the secondary critical field, the fluctuations are suppressed leading to a negative MR similar to a pure FM.

%is seen below the critical field which correlates to reordering of the spins to a ferromagnetic-like state which then is a negative magnetoresistance.  

In conclusion, the MR measurements along each crystallographic axis gives more insight on the unique character of the Fe$_3$Ga$_4$ phases. At high fields beyond saturation, the MR matches with the electron-magnon and cyclotronic model proposed by Khosla-Fischer with the Sondheimer-Wilson correction for the exponent. At low fields, a persistent positive MR along the $b$-axis emerges with a uncharacteristically high mobility at low fields, which indicates helical ordering instead of traditional FM ordering as originally expected with possible topological phenomenon. Future measurements may give insight on whether the topological hall effect (THE) could be found in this compound.

%Below this field, the MR is positive and increasing, which fits well with the hopping model proposed by Shklovskii and Efros

%\begin{equation}
%   ln\frac{\rho_B}{\rho_0} \propto \frac{B}{B_0}^m
%\end{equation}

%where $\rho_B$ is the resistivity with applied field, $\rho_0$ is resistivity with zero applied field, $B$ is the applied field, and $B_0$ is a constant, and m $\approx$ 2 for the hopping model. However, it is noted that 1 $<$ m $<$ 2 for a variety of systems\cite{saha}, including FeGe thin films with an intrinsic Berry phase mechanism\cite{Marrows2} ferromagnetic thin films and doped semiconductors. 

\begin{acknowledgements}
Research at the United States Naval Academy was supported by the NSF DMR-EPM 1904446, Kinnear Fellowship, and Office of Naval Research under Contract No. N0001423WX02132. Work at Northeastern University was partially supported by the National Science Foundation grant DMR-1905662 and the Air Force Office of Scientific Research award FA9550-20-1-0247 (D.H.). Work at VCU was partially funded by National Science Foundation, Award Number: 1726617. This work was also supported by the Office of Naval Research through the NRL basic research program. The National High Magnetic Field Laboratory is supported by the National Science Foundation through NSF/DMR-2128556 and DMR-1644779 and the State of Florida.
\end{acknowledgements}
\newpage
\label{References}

\end{document}

\begin{center}

\textbf{Supplementary Material for\\ ``Anomalous low-field magnetoresistance in Fe$_3$Ga$_4$ single crystals"}

\end{center}

\section{Crystallographic Information}
\begin{figure}[!h]
    \includegraphics[width=0.4\textwidth]{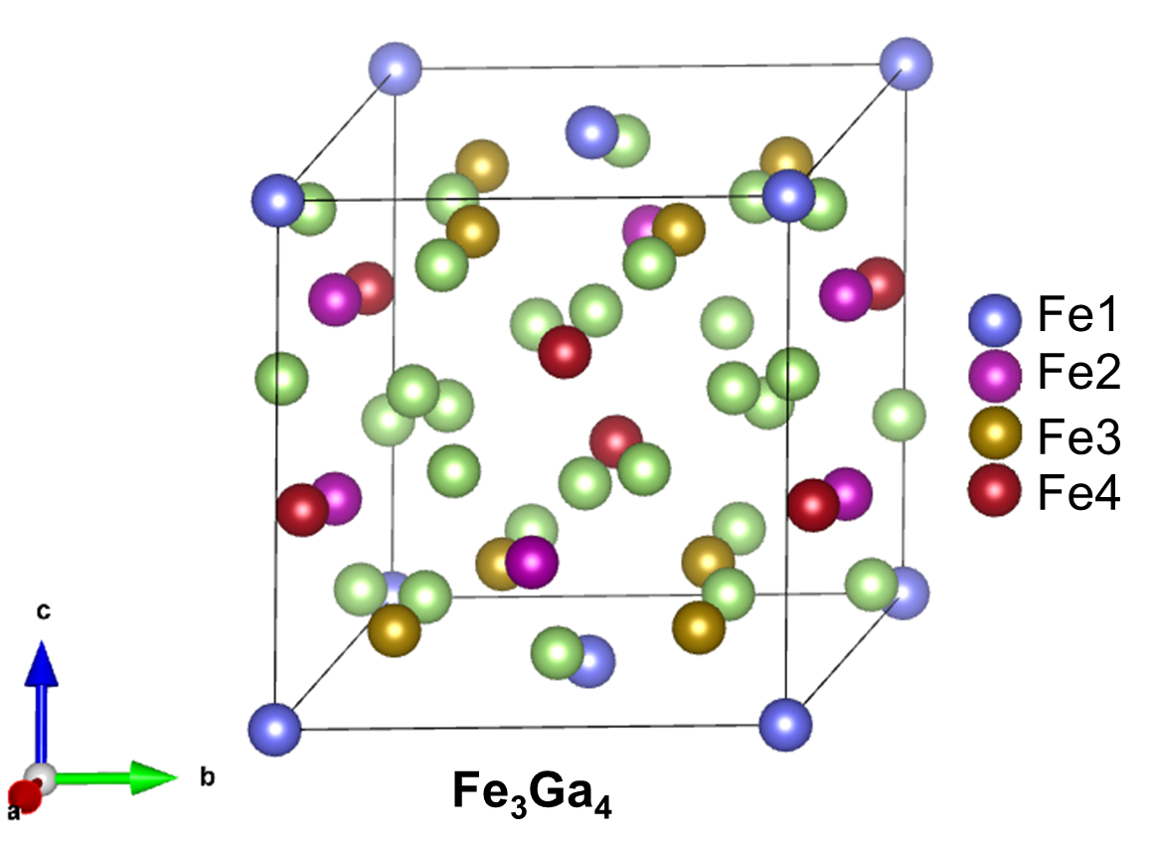}
    %\label{fig:crystal}
    \label{fig:crystal}
        \caption{The C2/m crystal structure of Fe$_3$Ga$_4$. The crystallographic properties of Fe$_3$Ga$_4$ in the monoclinic structure.}
\end{figure}
The structure forms in the C2/m crystal structure illustrated in Fig.\ref{fig:crystal}. The crystallographic positions of the Fe and Ga atoms are noted in the table portion of Fig. \ref{fig:crystal}. The structural properties of the crystals have been studied extensively in Wilfong et al. \cite{WilfongSC}
\begin{figure*}[!h]
\centering
\includegraphics[scale=0.3]{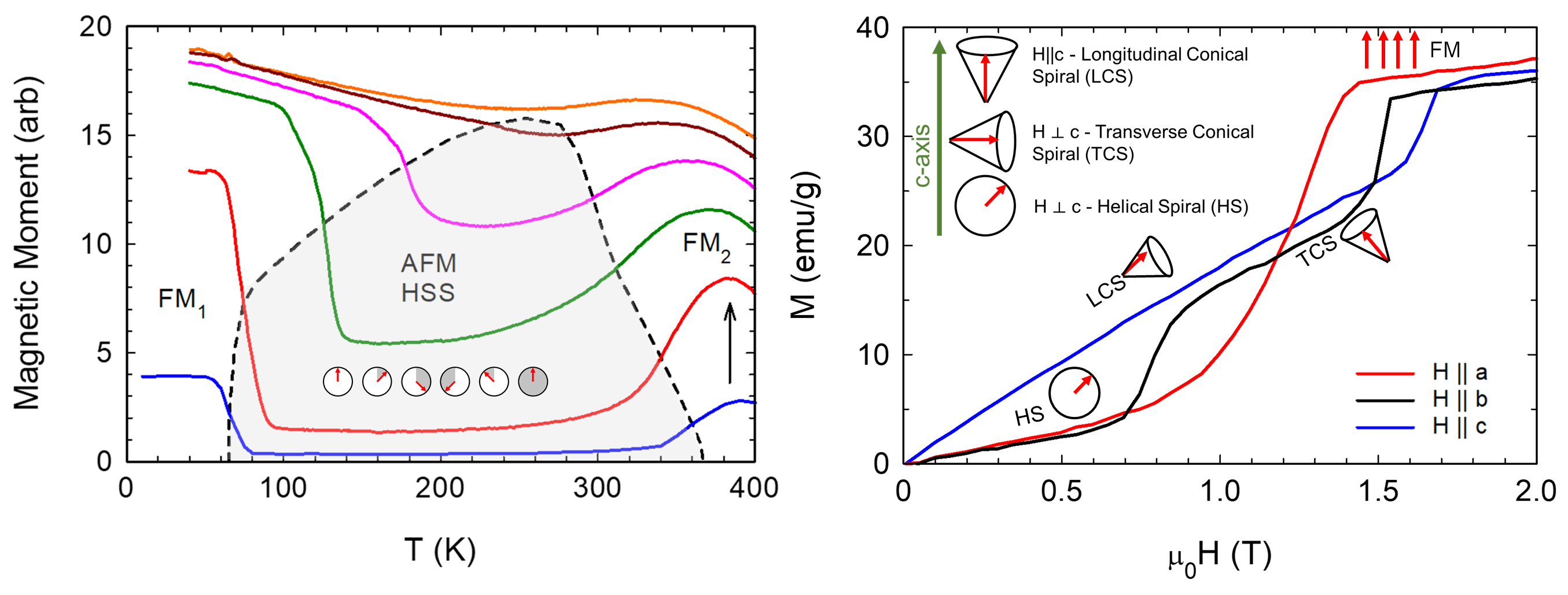}
  \label{fig:oldfig1}
    \caption{(a) The magnetization versus temperature measurements at fields between 0.2 - 3 T, with curves offset for clarity. The grey area denotes the helical spin spiral (HSS) or conical regions where the material reorders the spins from FM to AFM. (b) The magnetization versus field at 210 K along each crystallographic direction. The material is in the AFM state at the temperature, and the helical ordering is changing with applied field. Regions of spin alignment are noted. }
\end{figure*}

\newpage
\section{Magnetic properties}

The magnetic properties have been extensively studied for Fe$_3$Ga$_4$. The overview of the magnetic properties is seen in Fig. \ref{fig:oldfig1}. The average polycrystalline properties magnetization versus temperature, where the AFM-HSS data is highlighted in grey in Fig. \ref{fig:oldfig1}(a). The magnetization versus field data at T =140 K, where the crystal is in the HSS state is shown in Fig. \ref{fig:oldfig1}(b). The magnetic state is noted as helical spin (HS) along the $a$- and $b$-axis at low fields below 0.75 T. The tranverse conical spin is along the $a$- and $b$- axis above 0.75 T and below saturation. The longitudinal conical spin state is only observed along the $c$-axis.

\begin{figure*}[!t]
    \centering
    \includegraphics[scale = 0.8]{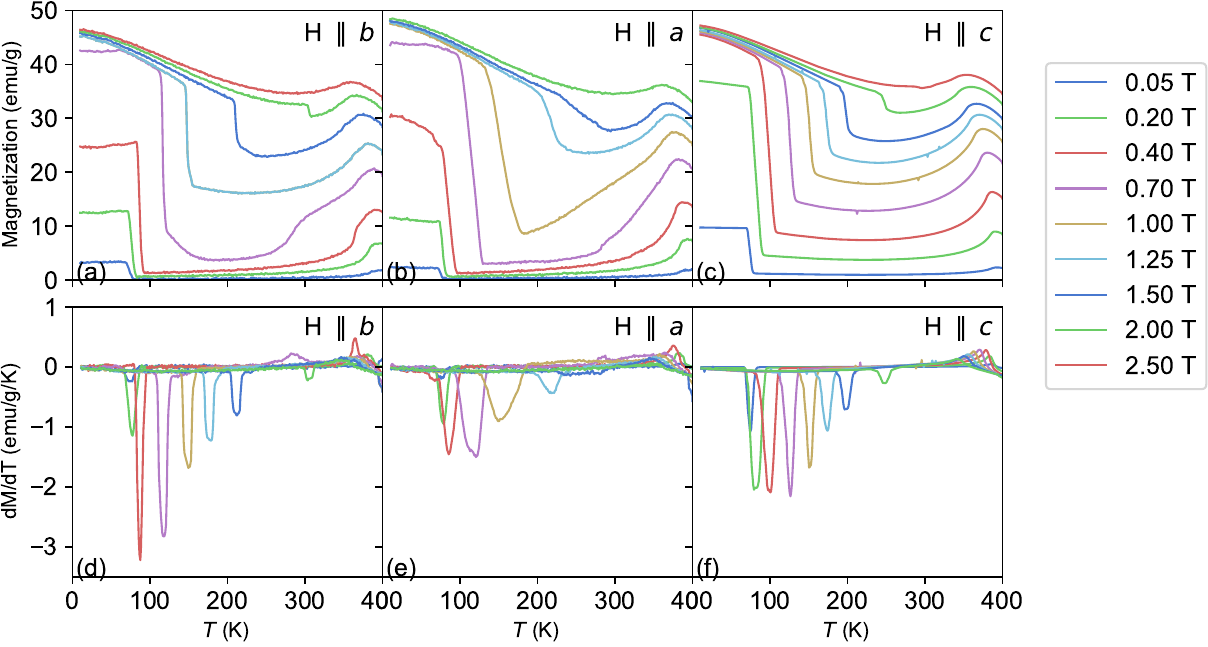}
    \caption{The magnetization versus temperature properties of the Fe$_3$Ga$_4$ single crystal with applied field parallel respect to the (a) b-axis, (b) a-axis, and (c)-axis direction. The derivative of the magnetization with respect to temperature (dM/dT) is also shown for each curve with respect to the (d) b-axis, (e) a-axis, and (f) c-axis. The dM/dT curve illustrates the changes in the magnetic transition temperature with applied field. As seen in both the M versus T and dM/dT versus T, the magnetic transition temperature is increasing with the applied field.} 
    \label{fig:magnetometry}
\end{figure*}

\begin{figure*}[!h]
\centering
\includegraphics[width=0.65\textwidth]{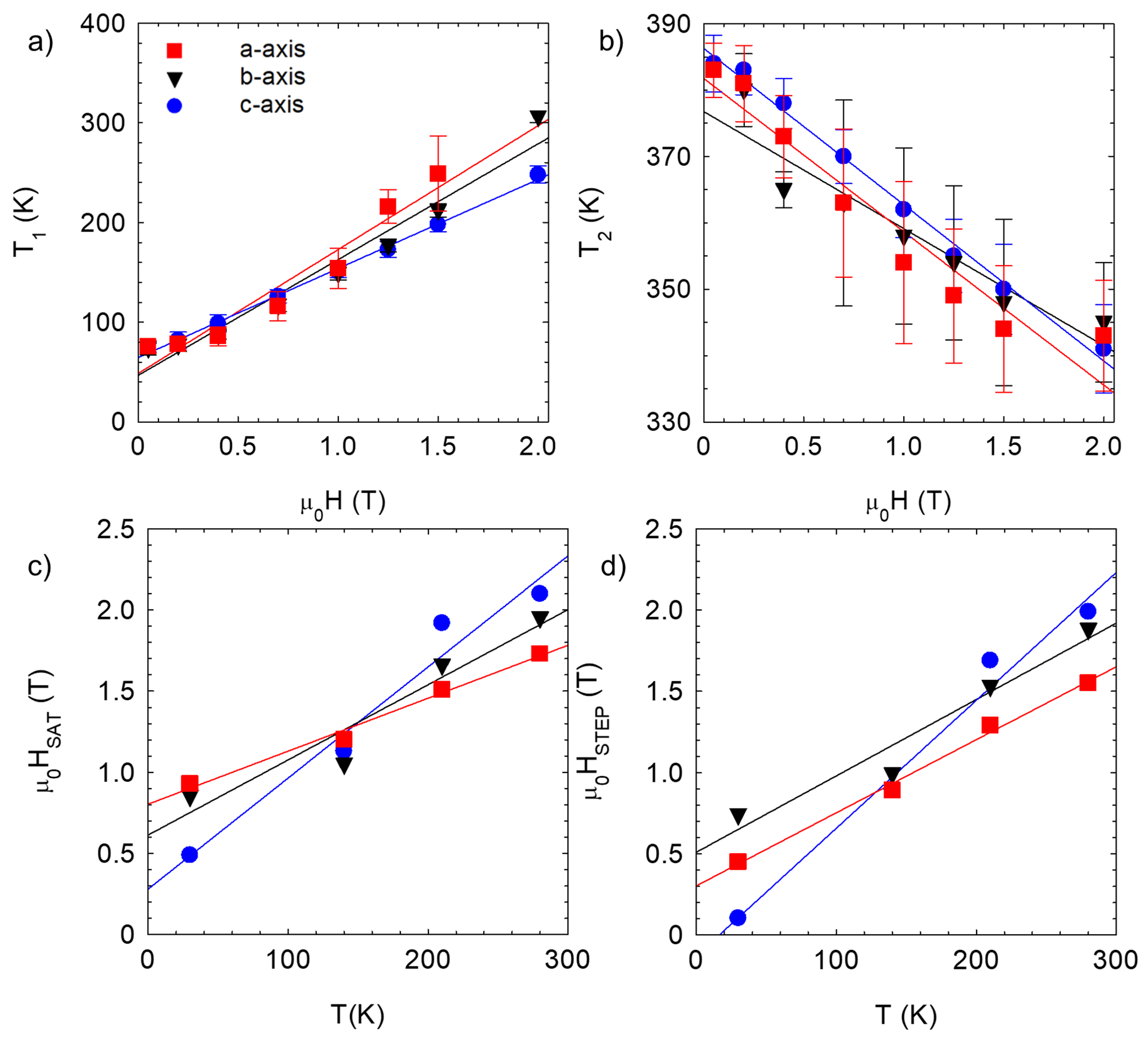}
  \caption{a) The metamagnetic transition as a function of temperature versus applied field. As noted in the figure, T$_1$ incereases linearly with applied field. b) The second transition T$_2$ as a function of temperature versus applied field. The temperature transition T$_2$ decreases with applied field. c) The saturation field ($\mu_0 H_{sat}$ as a function of temperature. As seen, the field where saturation occurs increases with temperature. d) The magnetic step where the spins reorient as a function of temperature. The steps correlate to the derivative as seen in \ref{fig:magnetometry}. Not shown is a consistent step at $\mu_0 H \approx 0.75$ T for the $b$-axis, which is shown in the main manuscript.}
    \label{fig:transitions}
\end{figure*}

The data for the magnetization versus temperature for each crystallographic direction is seen in Fig. \ref{fig:magnetometry}. The results for the temperature and field where the magnetic transitions T$_1$ and T$_2$ occur are plotted in \ref{fig:transitions}(a) and (b). The reported values of T$_1$ and T$_2$ were found by the peak location in the dM/dT, and the error is determined by the full width at half maximum. 

The magnetic step and magnetic saturation value was found by taking the derivative of the dM/dH curves, to find the exact field where there is a change in the magnetic ordering. As seen in Fig. \ref{fig:transitions}(c), the saturation field increases linearly with temperature. The magnetic step is where the spins change from FM ordering to spiral ordering or vice versa and noted in Fig. \ref{fig:transitions}(d). The magnetic step correlates to a decrease in the resistance in the MR, which then increases again when the field is at saturation.


\newpage
\section{Resistance Data}

\begin{figure}[!b]
    \centering
    \includegraphics[width=0.5\textwidth]{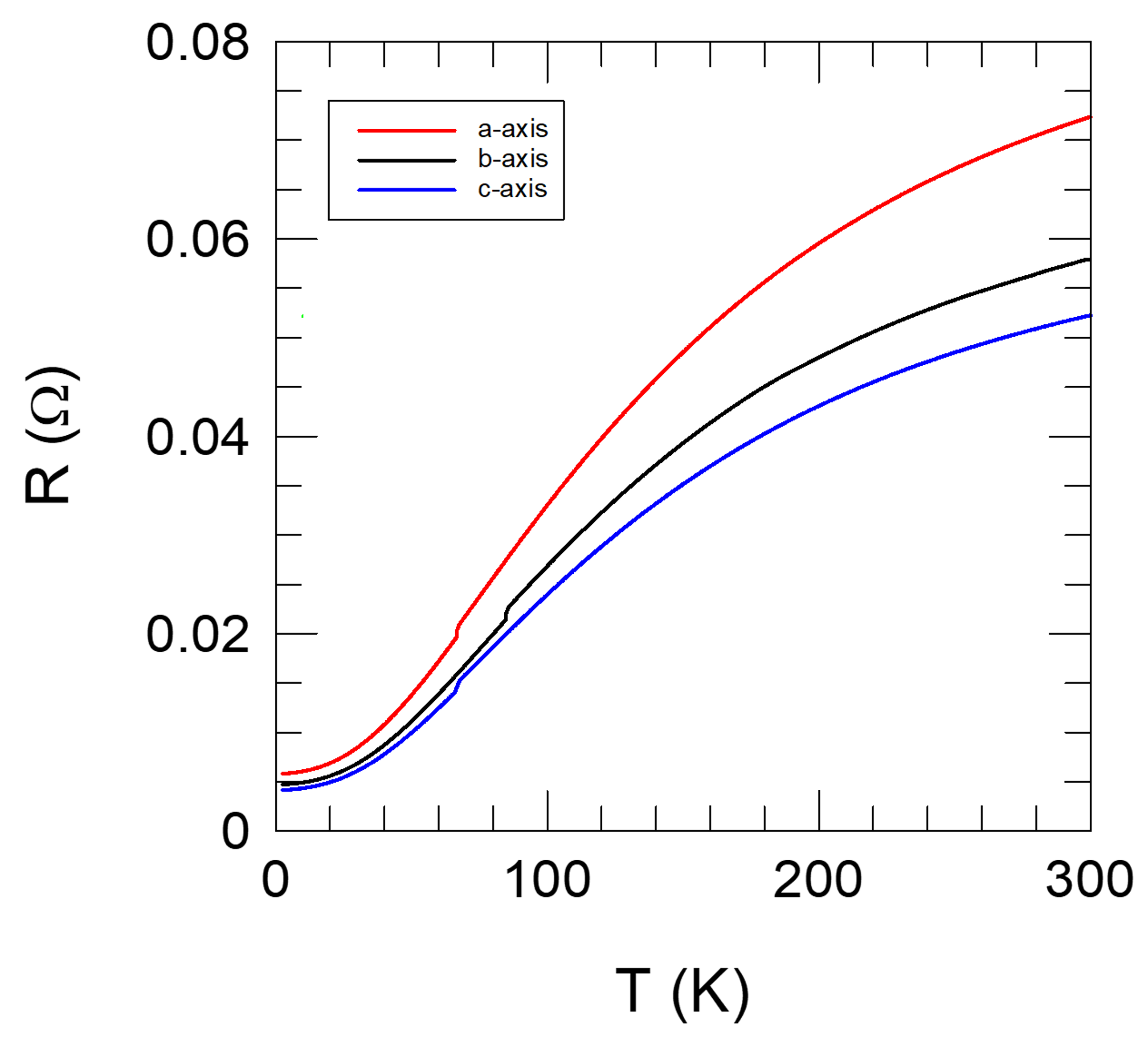}
    \caption{The $R(T)$ of the various axes with no applied field. All data is related to the cooling curve from 300 K - 2.5 K.}
    \label{fig:resistivity}
\end{figure}

The resistance data without applied field is seen in Fig. \ref{fig:resistivity} cooling from 300 K - 2.5 K. The data shows that there is change in the resistance at various temperatures, which is related to the spin reorientation from a FM to a HSS. 

\newpage
\section{Magnetotransport Data}

All fitting used a least-squares model to minimize $\chi^2$. The equation for the MR data in the medium and high-field regime is found by using the fit first used by Khosla and Fischer.
\begin{figure*}
    \centering
    \includegraphics[width=0.95\textwidth]{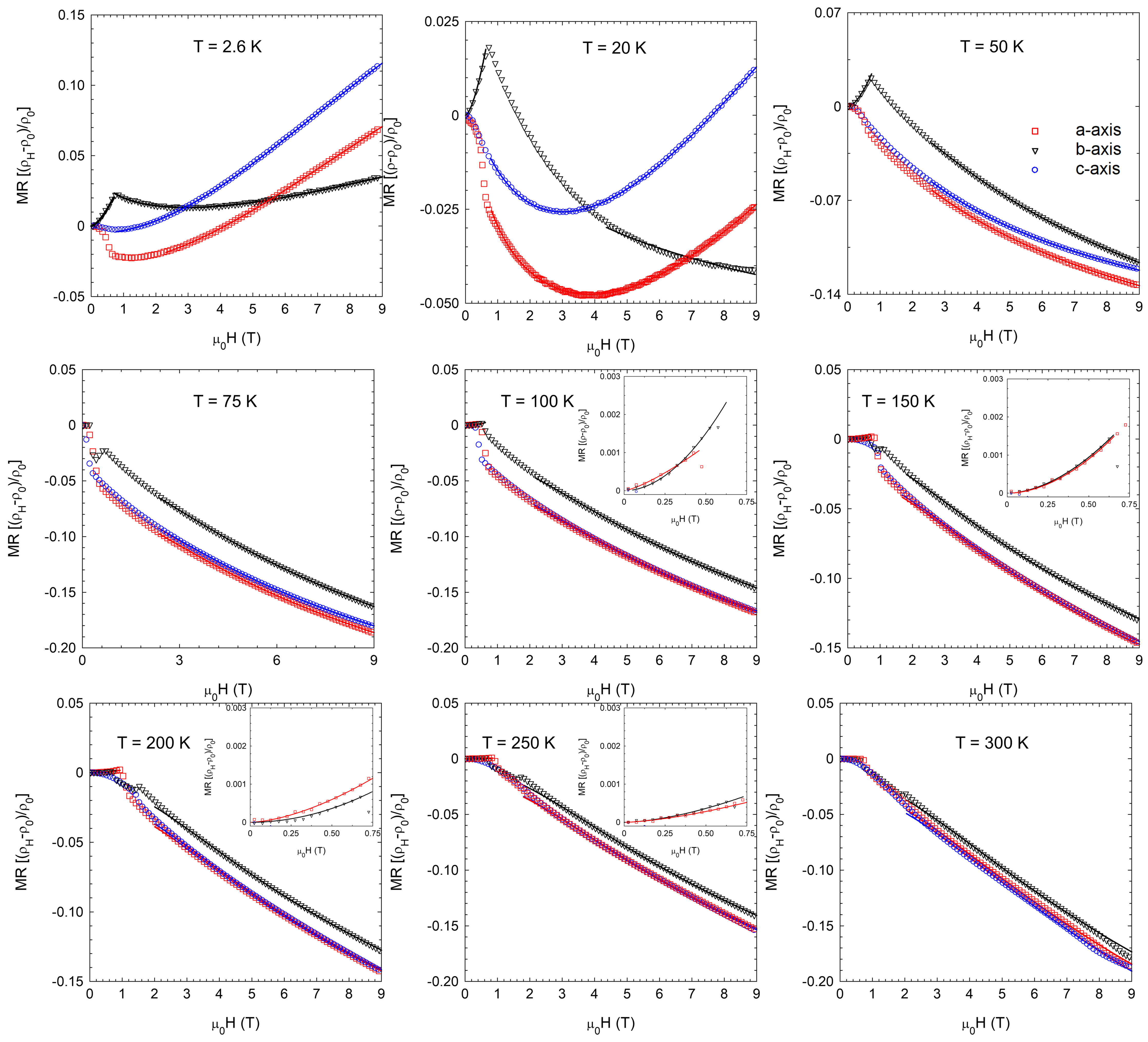}
    \caption{The magnetoresistance data for all temperatures plotted as a function of temperature for the $a$, $b$, and $c$ axes. The data above the FM to HSS (T $\leq$ 75 K) has the low-field positive MR data as an inset. The crystallographic axes are labeled the same for all data plots. The data with a straight line is the model shown in Table \ref{tab:fittingparameter}. For ease of viewing, the number of data points have been reduced.}
    \label{fig:MRdata}
\end{figure*}

\begin{equation}
\frac{\rho_H - \rho_0}{\rho_0} = -b_1 ln(1 + b_2^2 (\mu_0H)^2) + \frac{b_3^2(\mu_0H)^2}{(1 + b_4^2(\mu_0H)^2)}
\end{equation}

The values that are noted in the main manuscript are plotted, and these values are found in Table \ref{tab:fittingparameter}. 

\begin{table*}
\begin{center}

\scalebox{0.70}{
\begin{tabular}{ |p{1.75cm}|p{1.75cm}|p{1.75cm}|p{1.75cm}||p{1.75cm}|p{1.75cm}|p{1.75cm}||p{1.75cm}|p{1.75cm}|p{1.75cm}||p{1.75cm}|p{1.75cm}|p{1.75cm}|p{1.75cm}|p{1.75cm}||}
 \hline
 \multicolumn{13}{|c|}{Fitting Coefficients} \\
 \hline
 T (K) & \multicolumn{3}{|c||} {b$_1$} &\multicolumn{3}{|c||} {b$_2$$(\frac{cm^2}{Vs})$}&\multicolumn{3}{|c||} {b$_3$ $(\frac{cm^2}{Vs})$}&\multicolumn{3}{|c||} {b$_4$ $(\frac{cm^2}{Vs})$}\\
 \hline
  & a & b& c & a&b&c & a&b&c &a &b&c\\
  \hline
2.6                   &            &           &            &         &         &         & 1.82(2) & 0.311(2) & 23.6(5) & 78.1(4) & 0.00513(3) & 84.5(5) \\
20                    & 0.0108(11) & 0.0082(9) & 0.00986(2) & 3940(2) & 1220(1) & 2000(1) & 27.8(2) &          & 33.7(3) & 45.8(4) &            & 62.5(3) \\
50                    & 0.0325(4)  & 0.0457(5) & 0.0281(3)  & 618(3)  & 424(2)  & 926(5)  &         &          &         &         &            &         \\
75                    & 0.0471(6)  & 0.0538(6) & 0.0455(5)  & 355(2)  & 332(2)  & 370(2)  &         &          &         &         &            &         \\
100                   & 0.0528(6)  & 0.0555(6) & 0.0521(6)  & 308(2)  & 305(2)  & 315(2)  &         &          &         &         &            &         \\
150                   & 0.0539(7)  & 0.0539(7) & 0.0532(6)  & 316(1)  & 314(1)  & 321(1)  &         &          &         &         &            &         \\
200                   & 0.0602(7)  & 0.0595(7) & 0.0602(7)  & 280(1)  & 282(1)  & 284(2)  &         &          &         &         &            &         \\
250                   & 0.0687(8)  & 0.0692(8) & 0.0691(7)  & 269(1)  & 264(1)  & 273(1)  &         &          &         &         &            &         \\
300                   & 0.092(1)   & 0.0898(9) & 0.0981(1)  & 237(1)  & 240(1)  & 226(1)  &         &          &         &         &            &       \\ 

 \hline
\end{tabular}}
\label{tab:fittingparameter}
\caption{The fitting parameters for the medium and high-field portion of the magnetoresistance data.}
\end{center}
\end{table*}

The positive low-field components for the data were fit using another model that also varies as the field squared. While the $(\mu_0 H)^2$ component of the Khosla-Fischer model with the Sondheimer-Wilson correction could have also been used, this model was used to further highlight the interesting behavior of the low-field regime.

\begin{equation}
   \frac{\rho_H - \rho_0}{\rho_0} = (\mu \mu_0 H)^q
\end{equation}

The values for $\mu$ and $q$ are found in the Table \ref{tab:cycle}. All values are plotted in the main manuscript.
\begin{table}[]
    \centering
    \begin{tabular}{|p{3cm}|p{3cm}|p{3cm}|p{3cm}|p{3cm}|}
    \hline
     & \multicolumn{2}{|c|}{$\mu$} & \multicolumn{2}{|c|}{q}\\
     \hline
     T (K) & a $(\frac{cm^2}{Vs})$ &b $(\frac{cm^2}{Vs})$ & a   & b  \\
     \hline
     2.6    & - & 177(1) & - & 1.55(10)\\
     \hline
     20    & - & 125(1) & - & 1.59(05)\\
     \hline
    50    & - & 149(1) & - & 1.68(09)\\ 
    \hline
    75   & - & - & - & -\\
    \hline
    100 & 12.1(6) & 69.0(5) & 1.32(10) & 1.93\\
    \hline
    150 &47.9(5) & 42.9(5) & 1.88(02) & 1.81(02)\\
    \hline
    200 & 20.5(3) & 36.6(4) & 1.62(05) & 1.98(03)\\
    \hline
    250 & 8.96(10) & 15.1(4) & 1.62(05)& 1.51(05) \\
    \hline
    300 & - & - & -& -\\
    \hline
    \end{tabular}
    \caption{The values of the fit to the positive magnetoresistance at low field based on equation (2).}
    \label{tab:cycle}
\end{table}
\newpage
\section{Derivative of Magnetotransport Data}

The derivative of the magnetotransport data from Fig. \ref{fig:MRdata}. The derivative plots were used to find inflection points relating to $\rho_{low}$ and $\rho_{high}$. Above T$_1$ and above the saturation field, the $a$, $b$, and $c$-axis data has the same slope as seen by the derivative, which means that each crystallographic axis saturates at the same field, agreeing with Fig. \ref{fig:transitions}(c). Along the $b$-axis, there is a noticeable decrease in the MR at varying inflection points, which is enchanced in the derivative view. The magnetic step field seen in Fig. \ref{fig:transitions}(d) corresponds to the field where there is a step in the MR, and the saturation field corresponds to when the resistance increases again.
\begin{figure*}
    \center
    \includegraphics[width = 0.95\textwidth]{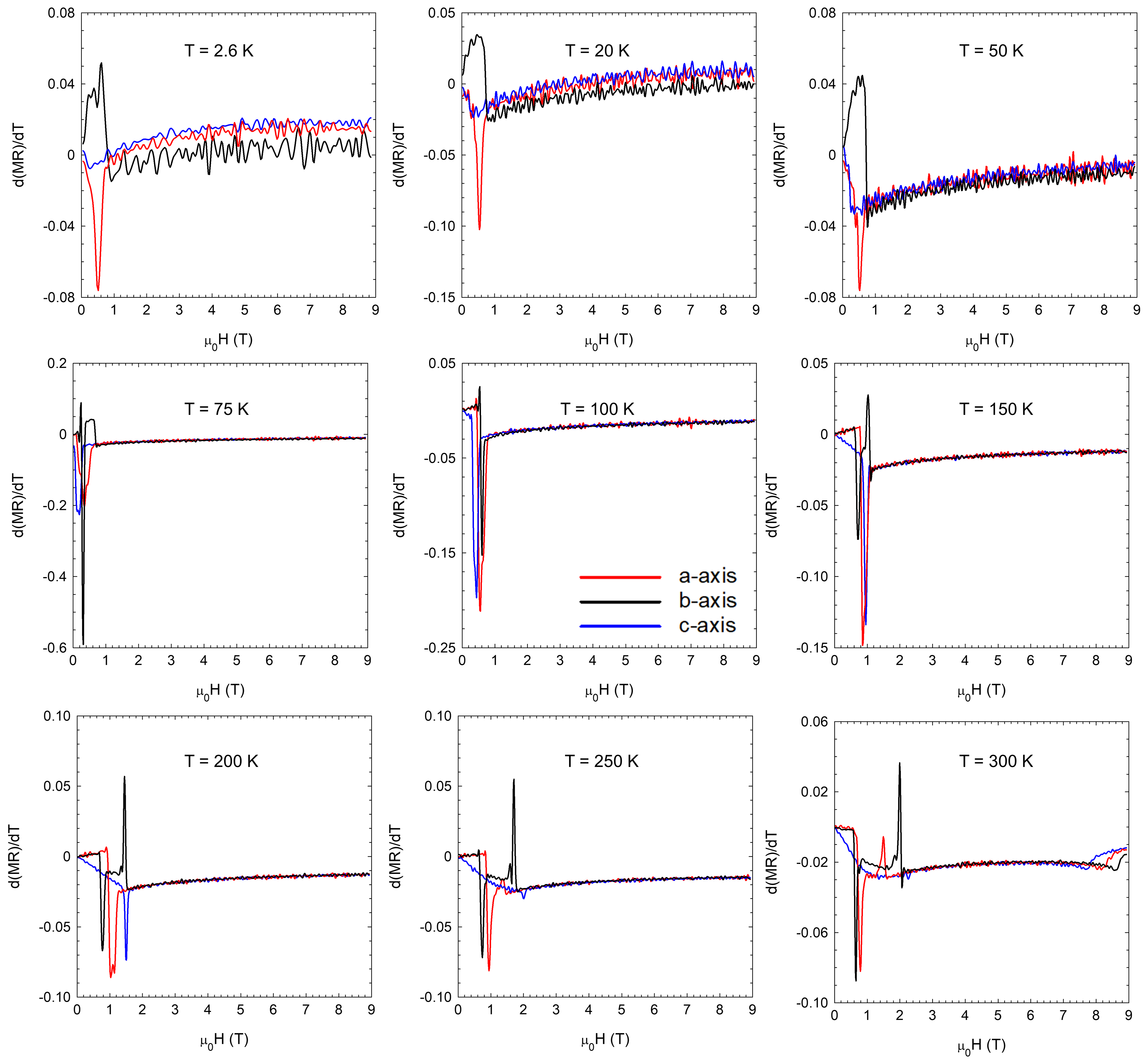}
    \caption{The derivative of the MR data with respect to field ($\frac{d(MR)}{dT}$). }
    \label{fig:derivative}
\end{figure*}

\clearpage
\label{References}